# GCC-Plugin for Automated Accelerator Generation and Integration on Hybrid FPGA-SoCs

Markus Vogt, Gerald Hempel, Jeronimo Castrillon
Technische Universität Dresden
Faculty of Computer Science
Dresden, Germany
Email: {forename.surname}@tu-dresden.de

Christian Hochberger
Computer Systems Group
Technische Universität Darmstadt
Darmstadt, Germany
Email: hochberger@rs.tu-darmstadt.de

*Abstract*—In recent years, architectures combining a reconfigurable fabric and a general purpose processor on a single chip became increasingly popular. Such hybrid architectures allow extending embedded software with application specific hardware accelerators to improve performance and/or energy efficiency. Aiding system designers and programmers at handling the complexity of the required process of hardware/software (HW/SW) partitioning is an important issue. Current methods are often restricted, either to bare-metal systems, to subsets of mainstream programming languages, or require special coding guidelines, e.g., via annotations. These restrictions still represent a high entry barrier for the wider community of programmers that new hybrid architectures are intended for. In this paper we revisit HW/SW partitioning and present a seamless programming flow for unrestricted, legacy C code. It consists of a retargetable GCC plugin that automatically identifies code sections for hardware acceleration and generates code accordingly. The proposed workflow was evaluated on the Xilinx Zynq platform using unmodified code from an embedded benchmark suite.

## I. INTRODUCTION

Today, embedded hybrid platforms combining field programmable gate arrays (FPGA) and high performance RISC processing cores give the user the freedom to implement specialized peripherals in the FPGA fabric while still relying on the execution power of the RISC processor(s). The Xilinx Zynq system on chip (SoC) family and the Altera Cyclone/Arria V SoC are prominent examples for this approach.

Such devices pave the path for the integration of arbitrary hardware accelerators in complex applications, however, most software developers are not familiar with hardware description languages (HDL). Thus, they are unable to develop application specific accelerators on their own. This problem has been addressed in the past by many researchers. Yet, the proposed solutions are not satisfactory. The user still has to write his own HDL code, has to take care of the HW/SW partitioning (often by annotating the existing code) and has to create the required SW/HW interfaces.

Our approach aims to notably lower the entry barrier for software developers to hardware-accelerated program execution. This particularly means using plain unannotated C, which is a popular and established language, as input. In this way, we bring hardware acceleration to a broader range of general applications. We envision a transparent workflow ideally not demanding any HDL skills or knowledge about the underlying hardware platform from the developer, providing seamless integration with the software environment.

The contributions of this paper are:
- Automated HW/SW partitioning using a GCC-plugin that extracts accelerators from C code and generates synthesizable HDL code.
- Automated and platform agnostic code patching enables seamless integration with software environment. Accelerator invocation remains completely transparent with optional fall-back to software execution.
- Support for legacy application code without annotations.

The rest of this paper is organized as follows. Section II presents the related work. In Section III we introduce the target hardware platforms of our proposed workflow. Section IV describes our workflow, the compiler plugin and its integration into GCC. Sections V and VI present the evaluation of our approach and discuss results. The remaining Sections draw conclusions and point out future work.

## II. RELATED WORK

Since the emergence of FPGAs, many efforts have been made to exploit the performance gain offered by reconfigurable logic with customized hardware accelerators. This especially holds true for hybrid FPGA architectures tightly coupling a general purpose processor with reconfigurable logic.

The most obvious, flexible but also the most challenging way is to write accelerators by hand using an HDL and manually perform all required integration with the software environment. An example is shown in [1]. Designing accelerator-based systems that way, requires strong skills in HDL as well as deep knowledge of the underlying hardware platform. The development process usually is time consuming and error-prone. Hence, the ability to implement such systems is left to the relatively small community of FPGA developers.

A number of approaches have been presented that reduce or even completely eliminate the necessity of writing HDL. The goal is to generate synthesizable code for accelerators from a more abstract problem description. LegUp [2] is an open source high-level synthesis (HLS) tool for FPGA based hybrid systems. The HW/SW partitioning is determined by profiling the C program on a self-profiling processor and altering the software binary afterwards in order to run it on the hybrid system. In [3] the authors present basic support for ARM-FPGA hybrid SoCs. In [4] the authors present Nymble, a system based on the techniques introduced by COMRADE [5]. It allows a much larger scope for accelerators by supporting a mechanism for back-delegation of unsuitable code sections into software. For HW/SW partitioning, Nymble requires additional code annotations using pragmas.

Nymble as well as LegUp use Low Level Virtual Machine (LLVM) as compiler framework. As shown in [6], the Gnu compiler collection (GCC) has been used for HLS workflows as well. The authors show a customized GCC compiler for generation of hardware accelerators for a bare-metal soft-core processor. Our work extends C-to-HDL transformations for better integration in more complex systems.

The Delft Workbench [7] is a toolset providing semi-automatic HW/SW partitioning as well as HLS for FPGA. The targeted Molen machine architecture can be regarded as hybrid FPGA-processor architecture. The candidate kernels for hardware acceleration are determined by profiling but must be extracted manually.

Xilinx provides Vivado [8], [9], one of the most popular commercial HLS tools. It supports translating C, SystemC or C++ code directly into hardware. Vivado aims at mapping the whole application to hardware, which requires manual HW/SW partitioning by the user. Similar to Vivado, other HLS tools like ROCCC [10] or CATAPULT [11] provide sophisticated hardware synthesis for hardware-only solutions, with no support for a hybrid HW/SW translation. In [12] the authors present a framework that matches portions of C code (algorithmic skeletons) exposing specific memory access patterns against a library of known accelerator templates. In [13] authors particularly address the integration of accelerators with the software domain. They present a linker that creates an executable by transparently linking functions implemented in software objects and/or hardware accelerators. With the runtime environment provided, programs can be executed on a Zynq platform running embedded Linux.

Most of the approaches mentioned so far address a certain task related to accelerator generation or integration, but the user still has to perform manual work. This requires, even though to a lesser extent, knowledge of HDL and the underlying hardware platform. In contrast, the work in [14] raises the level of abstraction to completely hide the HW/SW boundary from the software developer. The work in [14] applies the principle of binary acceleration, which means identifying sequences of processor instructions worthy of acceleration at runtime and migrating them to specialized execution units. However, live application analysis and accelerator synthesis typically require a reasonable amount of computational resources, pushing todays embedded runtime environments towards their limits.

A promising solution to the issues left open by the approaches mentioned above is to rethink the entire design flow. This has been done by the Liquid Metal project [15]. The authors developed the Lime language [16], enabling programmers to describe a system in a hardware-friendly but still object-oriented manner. Lime programs can be compiled either into pure software binaries or into software and a set of hardware accelerators. All interfacing is done automatically by the runtime environment. Introducing a well tailored language circumvents limitations that arise from using existing languages. However, adopting a new language is a high entry barrier for most programmers and existing software must be ported to benefit from hardware acceleration.

## III. PLATFORM

The work presented in this paper especially addresses recent hybrid platforms combining embedded processors with a reconfigurable fabric. In this section we briefly describe one such system, namely, the ZedBoard evaluation kit containing a Xilinx Zynq-7000 [17] device.

The programmable logic (PL) in the Zynq-7000 device is a full Artix-7 FPGA fabric, while the processing system (PS) is a complete ARM subsystem featuring a Cortex-A9 dual core processor and a comprehensive set of peripherals. The PS provides four 32-bit general purpose (GP) AXI interfaces, which allow connecting peripherals from the PL as well as four full-duplex 64-bit high performance (HP) interfaces for connecting AXI masters residing in the PL. The Zynq architecture provides one special high performance interface connected to the Accelerator Coherency Port (ACP). The ACP is internally connected to the ARM Snoop Control Unit and can be used for cache coherent accesses to the ARM subsystem.

It should be noted, that the specific handling of these different AXI interfaces depends on the hardware residing in the PL which presumes a profound understanding of the hardware accelerator.

## IV. WORKFLOW

The workflow for transparent HW/SW partitioning and compilation is composed of four steps as shown in Figure 1. (1) *loop data collection* performs a whole-program analysis collecting information about all loops across all compilation units. (2) *loop analysis* uses that information to select loops for potential HW acceleration, using a cost model of the target platform. (3) *hardware generation* performs an HLS of the loops selected by the previous step and (4) *application modification* adapts the original software code to integrate the accelerators and finally generates the application binaries.

Before discussing the steps in detail in Sections IV-B through IV-E, we briefly describe the compiler framework that was used for the workflow.

### A. Compiler Framework and Integration

The workflow in Figure 1 was implemented as two plugins for the GCC C compiler. GCC is one of the most widely used compilers for software development for embedded systems. Using such a mature and widely-used compiler framework helps to provide a full transparent workflow for software programmers.

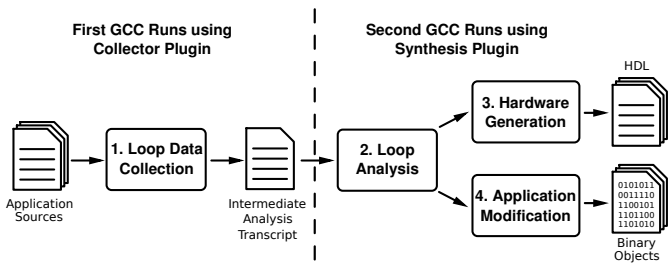

Figure 1. Abstract workflow for automatic accelerator generation

```
1  int fun3(int a, int b);
2  int fun1(int a, int b) {
3      int c;
4      for (int i=0; i<10; i++)
5          c += fun2(b + a, a - b);
6      return c;
7  }
8  int fun2(int a, int b) {
9      for (int i=0; i<30; i++) {
10         a += fun3(a, b);
11         b -= a;
12     }
13     return a+b;
14 }
```
Listing 1. Source code of **unit1.c**

```
1  int fun3(int a, int b) {
2      for (int i=0; i<100; i++) {
3          a += b;
4          if (a > 200 )
5              break;
6          b--;
7      }
8      return a;
9  }
```
Listing 2. Source code of **unit2.c**

GCC follows the traditional compiler structure divided into front-end, middle-end and back-end. Our analysis and transformations are performed in the middle-end on GIMPLE, GCCs internal intermediate representation. GIMPLE basically is a control flow graph (CFG) organized in basic blocks (BB) each containing statements in static single assignment (SSA) form. GIMPLE is further transformed into GCCs internal register transfer language (RTL) which finally is used by the compiler back-end to generate target specific machine code. All internal processing in GCC is controlled by its pass manager, while a pass refers to a certain transformation applied to the internal representation of the current compilation unit. In order to implement the steps depicted in Figure 1, custom passes are inserted using the pass manager.

To reason about the benefits of implementing a certain accelerator, one requires a global view of the application. However, GCC processes each file as separate compilation unit, which hinders whole program analysis. To overcome this drawback, GCC provides a link time optimization (LTO) framework which enables assorted optimizations during link time by storing the GIMPLE representation of each translation unit in the associated object file. Unfortunately, LTO only provides limited hooks for custom passes. We enable whole program analysis without using LTO by running the compilation flow twice (left and right part of Figure 1). The first run collects all information providing the second run with an overall view of the whole application. This global view is required in order to find accelerator candidates. The two consecutive compiler runs are wrapped by GNU Make to remain transparent to the user.

Since version 4.5.0 GCC provides a plugin interface for custom optimization passes, which are invoked by the pass manager using callback functions. The passes described in the following sections are implemented as two plugins for GCC 4.8.3, the *collector plugin* and the *synthesis plugin*. As the plugins work with an existing compiler binary, building a cross compiler for the target architecture is not required. Nevertheless, the plugins themselves must be built for a specific target architecture and hardware interface. Currently, we support the ARM architecture with AXI bus interface and an additional FPGA-based soft-core processor [18] using its proprietary bus interface.

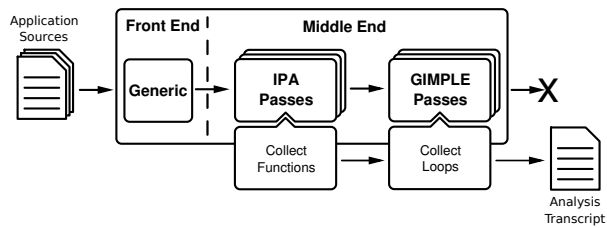

Figure 2. First GCC run invoking the *collector plugin*

### B. Loop Data Collection

The first GCC run invokes the *collector plugin* (step 1 in Figure 1), which implements the two custom passes *collect functions* (CF) and *collect loops* (CL) as shown in Figure 2. The CF pass is executed after all the inter-procedural passes (IPAs) have run. At this point, the compiler knows all functions declared and called in the translation unit. This information is preserved for later use. The CL pass runs after the GIMPLE loop optimizer passes, when all loops in the translation unit have been processed by the compiler. We now collect the compilers internal profiling data for each loop, which includes the local iteration count and a list of called functions and accessed memory locations. The data gathered by the passes CF and CL is accumulated in a single analysis transcript file.

Listing 3 shows a simplified version of the analysis transcript file after compiling the source files shown in Listings 1 and 2. The property `well_nested` indicates whether a loop or loop nest is synthesizable at all. Since loop 1 and 2 both include function calls, `fun2` and `fun3` respectively, only loop 3 will further be considered for accelerator generation.

### C. Loop Analysis

The second GCC run invokes the *synthesis plugin* (steps 2, 3 and 4 in Figure 1) which implements the custom pass

```
1  unit1.c                11    call=1
2   function=fun2         12    well_nested=0
3    loop1                13    -fun2
4     count=29            14
5     call=1              15 unit2.c
6     well_nested=0       16  function=fun3
7     -fun3               17   loop3
8   function=fun1         18    count=99
9    loop2                19    call=0
10    count=9             20    well_nested=1
```

Listing 3. Analysis transcript file after compiling **unit1.c** and **unit2.c**

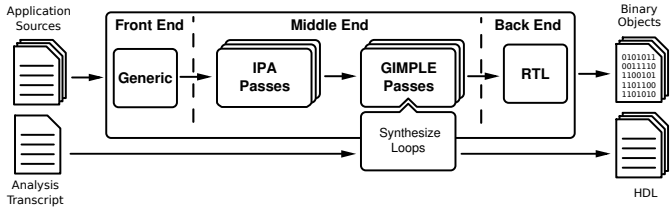

Figure 4. Second GCC run invoking the *synthesis plugin*

*synthesize loops* (SL) as shown in Figure 4. To ensure consistency of GCCs internal GIMPLE representation, the passes CL (first run) and SL (second run) must be invoked at the same processing stage within each run. In step 2 the SL pass reads the transcript file written by the collector plugin and constructs a call graph for the whole application. Based on this graph the total iteration count for each loop is estimated. Now, by default, all loops are ordered by their total iteration count in order to select the first $n$ loops or loop nests as synthesis candidates. The value for $n$ is a runtime parameter for the compiler plugin and the sorting function is customizable to consider other loop properties, e.g. instruction count. This enables the implementation of an arbitrary cost model to sort the loops.

### D. Hardware Generation

In step three of our workflow we generate an HDL implementation for all loops selected by the previous step. This is accomplished by translating the GIMPLE CFG of each loop or loop nest into a finite state machine (FSM). If this step discovers GIMPLE statements or operands which cannot be handled, a compiler warning is generated and the loop candidate is rejected.

During the generation of the FSM a number of optimization techniques are applied. Namely, speculative execution of conditional branches in parallel, list or modulo scheduling, and chaining of consecutive arithmetic operations. We do not explicitly address resource sharing, since FPGA vendor tools achieve better results for that purpose [19].

We are able to estimate the number of clock cycles for worst and best case execution as we know the clock cycle overhead of the accelerator invocation, the clock frequency ratio to the host processor and the shortest and longest path of our FSM. Furthermore, we define an architecture-specific penalty for memory accesses. Along with these heuristic, we are able to estimate the speedup of the accelerator in question. If the results do not meet the constraints specified as compilation parameters, the accelerator is rejected.

The final HDL implementation of the accelerator consists of a loop specific and a target specific part. The former implements a combination of FSM and datapath with a generic register and memory interface. The latter adopts this interface to a target specific host processor interface, e.g. a certain peripheral bus architecture. For example on Zynq, the accelerator is integrated as AXI peripheral module into the system.

### E. Application Modification

The final step of our workflow modifies the original code in order to call the synthesized accelerators from the application. We use an abstract calling scheme from the applications point of view. This decouples the code patching from the actual communication protocol. Therefore, each accelerator invocation is wrapped by a generated function. Its implementation is emitted as C code and provides input and output arguments for data transfer between application and accelerator. In our implementation, this function determines the base address of the called accelerator, writes input values to registers, starts the accelerator and reads back output values on return.

The call to that wrapper function is placed preceding the BB of the original loop header, as depicted in Figure 5. It shows the original and modified GIMPLE graphs of the loop in `fun3()` (Listing 2). The inserted variables `tmp.9` and `tmp.16` provide the return values from hardware. They correspond to the original loop exit variables `a.6` and `a.7` respectively. To properly retain control flow in case of multiple loop exits, the accelerator always returns `bb_idx`, which denotes the basic block of the original loop the exit condition occurred in. This value is evaluated by inserted conditional branches directing control flow to the corresponding BB after the original software loop. This bypass is inserted between the wrapper function call and the original loop header. Preserving the original loop enables fall-through to software execution. Resource allocation and sharing techniques can be then applied, since the application is still functional in case no accelerator is currently available.

The presented calling scheme requires inserting a few GIMPLE instructions only while providing the whole flexibility of C for implementing the actual hardware access. This further manifests when targeting platforms running a full-blown OS, such as Linux on the Zynq platform. Such systems require invocation of device drivers for accessing the underlying hardware, which could be rather complex. Furthermore, the wrapper function could be modified or extended by arbitrary user code with little effort. This especially enables debugging of the accelerator call using additional code or even breakpoints.

## V. EVALUATION

The evaluation presented in this section pursues three different goals, namely, (i) test the prototype of our seamless programming flow on a sample application, (ii) show the generality of our approach by applying it to arbitrary unmodified code, and (iii) analyze the challenges posed by today's hybrid platforms for automated accelerator generation (Section VI).

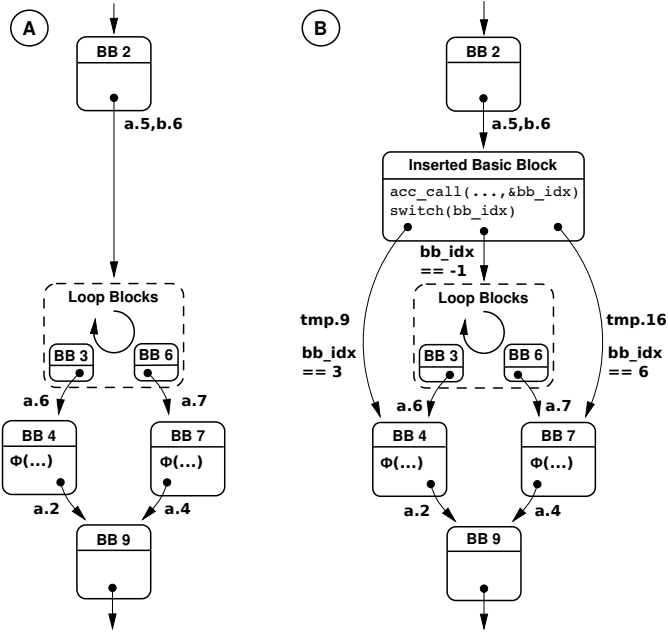

Figure 5. (A) Original GIMPLE Graph; (B) Modified GIMPLE graph with accelerator call

### A. Line of Sight

The operability of our approach was demonstrated with a 2D line-of-sight algorithm which determines whether a line intersects a square by iteratively checking each point on the line. Our GCC workflow transforms the main loop of this algorithm into an FSM with 10 states. The generated interface used 11 input registers and one output register in order to exchange data with the accelerator.

Our prototype was implemented on a ZedBoard running Arch Linux. The ARM processor was clocked at 666 MHz, while the accelerator operated at 333 MHz. Both components were connected via a GP AXI port. Using an HP port would not have improved performance, since we have not yet implemented accelerator controlled memory access. We tested our implementation using random input data to eliminate run time dependencies on the data. To prove correctness, we compared the results calculated in software with those calculated by the accelerator hardware. Table I shows the execution times of the accelerator and software-only version. The overhead of an accelerator call is $\approx 2.4\,\mu s$. This value is composed of a relatively small amount for the inserted software instructions and a larger amount for data transfer. This is due to the latency of 14 accelerator clock cycles for each AXI register read or write operation.

Further, we compared our results with LegUp HLS. As LegUp has special requirements in order to perform automated HW/SW partitioning, we run their HLS compiler stand-alone on the portion of C code that was identified as accelerator by our toolflow. The resulting FSM has 3 states and runs at a maximum speed of 170 MHz as indicated by Xilinx ISE syn-thesis results. This comparison shows that our HLS approach requires improvement but also that even established HLS tools hardly outperform the ARM processor. In Section VI it is discussed whether single problem speedup is required at all to gain overall system speedup.

Table I
EXAMPLE ACCELERATOR EXECUTION TIMES AND CALL OVERHEAD

|  | Maximum Clock Rate | FSM States | Execution Time | Relative Performance |
|---|---|---|---|---|
| Software | 666 MHz | – | $62\,\mu s$ | 1.00 |
| Hardware | 333 MHz | 10 | $126\,\mu s$ | 0.49 |
| LegUp[†] | 170 MHz | 3 | $74\,\mu s$ | 0.83 |

[†]Estimation based on synthesis results using Xilinx ISE

We could demonstrate the generation of a complete and correct working HW/SW implementation from plain C without user intervention. This example shows that patching on GIMPLE level is a viable approach for seamless accelerator integration.

### B. MiBench

In order to demonstrate the generality of our approach we compiled MiBench [20]. This embedded benchmark suite addresses real world problems and contains code from six different application domains. For this test we considered all accelerators found and did not apply any estimation of the expected speedup.

Due to limits in the current implementation, which are further discussed in Section VI, we could not implement and run the accelerated applications. However, the results in Table II clearly prove the generality of our approach. We are able to find and synthesize a reasonable number of accelerators from unmodified code of various application domains. Furthermore, our tests demonstrate the stability of the tool flow while analyzing a large, arbitrary codebase.

Table II
ACCELERATORS SYNTHESIZED FROM MIBENCH

| Application Domain | Benchmark/ Library | Application Accelerators | Translation Units |
|---|---|---|---|
| Network | patricia | 4 | 1 |
|  | dijkstra | 2 | 2 |
| Consumer | lame | 5 | 17 |
|  | jpeg | 79 | 49 |
|  | tiff-v3.5.49 | 149 | 44 |
|  | mad-0.14.2b | 44 | 35 |
| Office | libsphinx2 | 39 | 45 |
|  | ispell | 1 | 3 |
|  | stringsearch | 11 | 5 |
|  | ghostscript | 80 | 51 |
| Automotive | bitcount | 2 | 8 |
|  | basicmath | 1 | 5 |
| Telecomm. | FFT | 3 | 3 |
|  | gsm | 23 | 24 |
|  | CRC32 | 1 | 1 |
| Security | pgp | 107 | 42 |
|  | sha | 9 | 2 |
|  |  | Σ 550 | Σ 337 |

## VI. RESULTS DISCUSSION

With our results we have shown that, firstly, our approach is valid. It can generate a working application with an attached accelerator not requiring any user interaction. Secondly, we are able to find a reasonable number of accelerators in arbitrary, unmodified code of real-world applications.

Currently, our approach is limited by our basic HLS algorithm and missing support for multiple accelerators. The focus of our work is in the GCC-plugin infrastructure that allows to transparently compile C code. Advances in HLS algorithms can be either integrated later, or we could call an existing HLS solution from within the GCC infrastructure.

Supporting multiple accelerators is a precondition for the evaluation of complex application scenarios. Since our approach wraps accelerator calls by a C-function, we plan to implement a device driver that is able to handle an arbitrary number of accelerators.

While the limitations previously mentioned do not hinder basic evaluation, exploring complex application scenarios is not feasible yet. Particularly we were not able to run any of the MiBench test cases, even though the suite offers large potential for acceleration as shown in Table II.

On platforms like Zynq, in terms of speedup, one challenge remains: One has to generate hardware running on the FPGA that outperforms the highly optimized ARM core featuring, e.g., conditional instructions and out-of-order execution. We believe that increasing single-accelerator performance by improved HLS or higher clock rates is not the only way to gain overall application speedup. Instead, acceleration potentially could be achieved by increasing thread level parallelism rather than execution speed of a single task. This is a truism since processor vendors moved towards multicore architectures. Utilizing multiple accelerators simultaneously leads from pseudo-parallelism to real task level parallelism beyond the number of present CPU cores. Such a system-level solution may provide a speedup even with single accelerators running slower than the CPU. Furthermore, with dedicated accelerators, it is to expect that energy-efficiency would also increase, which has to be confirmed by future work.

## VII. CONCLUSION

We presented a GCC-based workflow for accelerator generation and integration. It performs automatic HW/SW partitioning by synthesizing frequently executed loops to HDL. For seamless interfacing, the program code is patched on an abstract level not depending on target platform or accelerator interface. The whole process is neither demanding HDL skills from the user nor requiring knowledge about the underlying platform. The proposed workflow has been validated by implementing a working example on a Zynq platform.

Furthermore, a complex codebase has been compiled to demonstrate the generality of our toolflow. More than 500 accelerators could be generated from the sources of MiBench [20]. This complex example was not evaluated on the hardware platform due to current limitations of our workflow mentioned in Section VI.

## VIII. CURRENT AND FUTURE WORK

To overcome current limitations we work on implementing memory access for accelerators using the ACP available on Zynq devices as well as on full integration of an arbitrary number of accelerators into the OS using a device driver. Finally we want to bundle generated hardware in form of bitfiles with the application binary. This allows instant loading and execution of any accelerated application.

Beyond that, future work addresses improvements in HLS, in particular integrating with existing HLS approaches, and the migration to partial reconfiguration. HLS improvements may include exploiting more GCC-internal optimizations by moving hardware generation after the last GIMPLE optimization pass.